\documentclass[preprintnumbers,amsmath,amssymb,twocolumn]{revtex4}
\usepackage{graphicx}
\usepackage{dcolumn}
\usepackage{bm}
\usepackage{natbib}
\usepackage[caption=false]{subfig}

\begin{document}

\title{Ramsey interferences and spin echoes from electron spins \\
inside a levitating macroscopic particle}

\author{T. Delord$^{1}$}
\author{P. Huillery$^{1}$}
\author{L. Schwab$^{1}$}
\author{L. Nicolas$^{1}$}
\author{L. Lecordier$^{1}$}
\author{G. H\'etet$^{1}$} 

\affiliation{$^1$Laboratoire Pierre Aigrain, Ecole normale sup\'erieure, PSL Research University, CNRS, Universit\'e Pierre et Marie Curie, Sorbonne Universit\'es, Universit\'e Paris Diderot, Sorbonne Paris-Cit\'e, 24 rue Lhomond, 75231 Paris Cedex 05, France.}

\begin{abstract}
We report observations of Ramsey interferences and spin echoes from electron spins inside a levitating macroscopic particle.  The experiment is realized using nitrogen-vacancy (NV) centers hosted in a micron-sized diamond stored in a Paul trap both under atmospheric conditions and under vacuum. Spin echoes are used to show that the Paul trap preserves the coherence time of the embedded electron spins for more than microseconds. Conversely, the NV spin is employed to demonstrate high angular stability of the diamond even under vacuum. These results are significant steps towards strong coupling of NV spins to the rotational mode of levitating diamonds. 
\end{abstract}
\maketitle

Being able to prepare arbitrary motional quantum states of massive oscillators will be an important step forward for modern quantum science \cite{aspelmeyer}. 
Tremendous efforts are made towards this goal with experimental platforms ranging from clamped nano-fabricated devices \cite{connel, Teufel, Hong203} to levitating objects \cite{Chang19012010, Romero2}.
One recently proposed way to engineer motional states consists in coupling atomic spins to the motion of macroscopic objects using magnetic fields \cite{Treutlein, Rabl, Arcizet, Kolkowitz, ma,yin, Kumar, delord2017strong}.  The idea is to exploit the point-like character of single atomic electrons and their magnetic sensitivity to detect and also to couple them to the motion of the oscillator.  The resulting spin-dependent force on the object could for instance be used to cool the oscillator motion down, to generate motional Schr${\rm \ddot o}$dinger cat states, or entangle the spin with the mechanical oscillator \cite{yin,yin2013optomechanics}.  The platform can then be employed for sensing the mechanical zero-point fluctuations \cite{Kolkowitz} and for more fundamental tests of quantum mechanics \cite{Scala, Wan, Bose2}.

Amongst the vast range of mechanical oscillators, particles levitating in harmonic potentials are being investigated widely. 
They have shown record high quality factors, stemming mostly from the absence of clamping losses \cite{Chang19012010, romero2010toward, Romero2}. They also enable efficient tuning of the mechanical properties.
Lowering the trap stiffness after cooling the centre of mass mode could for instance increase the ground state wave-function spread to several $\mu$m, offering prospects for optical manipulations of the wavepacket \cite{Kaltenbaek2016}. There has also been several experiments that achieved spin read-out of NV centers hosted in diamonds that are trapped both in liquid \cite{Geiselmann, Horowitz} or under vacuum~\cite{Hoang, Neukirch2, vacuumESR, Pettit}, which are important steps towards coupling spins to macroscopic particles' motion. The main quantum physics tools for engineering internal electronic and motional states, namely Ramsey interferometry and spin echoes, are however still elusive with trapped macroscopic particles, stemming mostly from the high degree of control required on the particle external and internal degrees of freedom.

\begin{figure}[ht!!]
\centerline{\scalebox{0.35}{\includegraphics{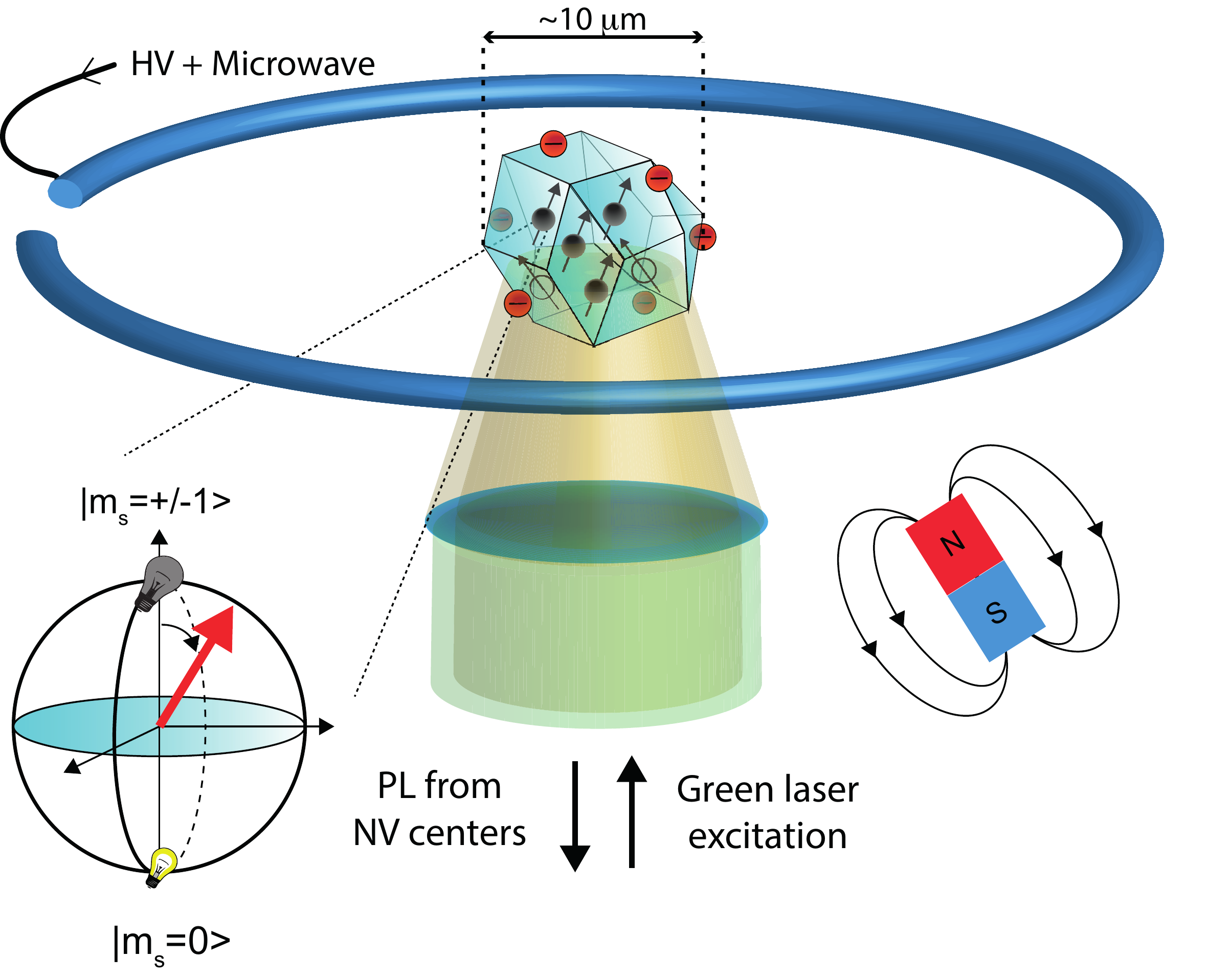}}}
\caption{Schematics of the experiment. A charged micron-sized diamond containing NV centers is levitating in a Paul trap. The spin properties of the NV centers are analyzed using confocal microscopy with both oscillating and static magnetic fields. The trap can be operated under vacuum or atmospheric conditions.}\label{setup}
\end{figure}

Ramsey interferometry and spin echoes have been the workhorse of many experiments in atomic physics for the past decades. They have allowed environmental noise control of the environment of electronic spins with atomic ensembles or single atoms for decades. Such measurements tools are also essential for prospective spin coupling to the motional state of macroscopic oscillators \cite{Kolkowitz, Bennett, yin2013optomechanics}. 
In this letter, we demonstrate contrasted Rabi, Ramsey oscillations as well as spin echoes from spins inside a macroscopic levitating particle under atmospheric conditions and under vacuum.
The experiment consists in manipulating the spin of NV centers within a micro-diamond levitating in a ring Paul trap. 
Although high voltages are used around small heavily charged diamonds, no differences between the longitudinal and transverse coherence times $T_1$ and $T_2^*$ of the NV spins inside and outside the trap are observed. Importantly, our experiments are performed in the presence of a large magnetic field, which demonstrates coherent control over angularly stable particles, an important prerequisite for recently proposed experiments on spin-mechanical coupling using the rotational modes \cite{delord2017strong, yin}.

Fig. \ref{setup} shows the principle of the experiment. A micron-sized gold-plated tungsten ring is used both for trapping micron-sized diamonds and for generating the oscillating transverse magnetic field that drives NV centers spins. 
A green laser is focussed onto the diamond and the photoluminescence (PL) from the embedded NV centers is collected using the same objective and directed to an avalanche photodiode. A permanent magnet is brought in the vicinity of the trap to lift the degeneracy between the $|m_s=\pm 1\rangle$ NV electronic spin states.
The NV centers are polarized using around 1 mW of green laser and a maximum of 20 dBm of microwave power can be brought to the ring to drive the electronic spins.
More details on the experiment, such as the number of NV centers or the diamond shape and size, are presented in \cite{delord2016, vacuumESR} and in the SI (section I). Compared to the experiment performed in \cite{vacuumESR}, the ring trap is 4 times smaller. We measured it to be 180 $\mu$m in diameter, which means that both the confinement and microwave powers at the trap center are greatly increased.  Using such a tiny ring trap is crucial for reaching angular stability under vacuum. 
%

Before realizing experiments with levitating particles, an in-depth study of the spin properties of NV from diamonds that are cast on a quartz plate was realized beforehand. It is presented in the SI (section II). Since at present we cannot compare the properties of the same particle in and out of the trap as was done in \cite{Kuhlicke}, such a characterization step was mandatory to estimate the influence of the trap on the NV photo-physical properties. 

\begin{figure}[ht!]
\centerline{\scalebox{0.28}{\includegraphics{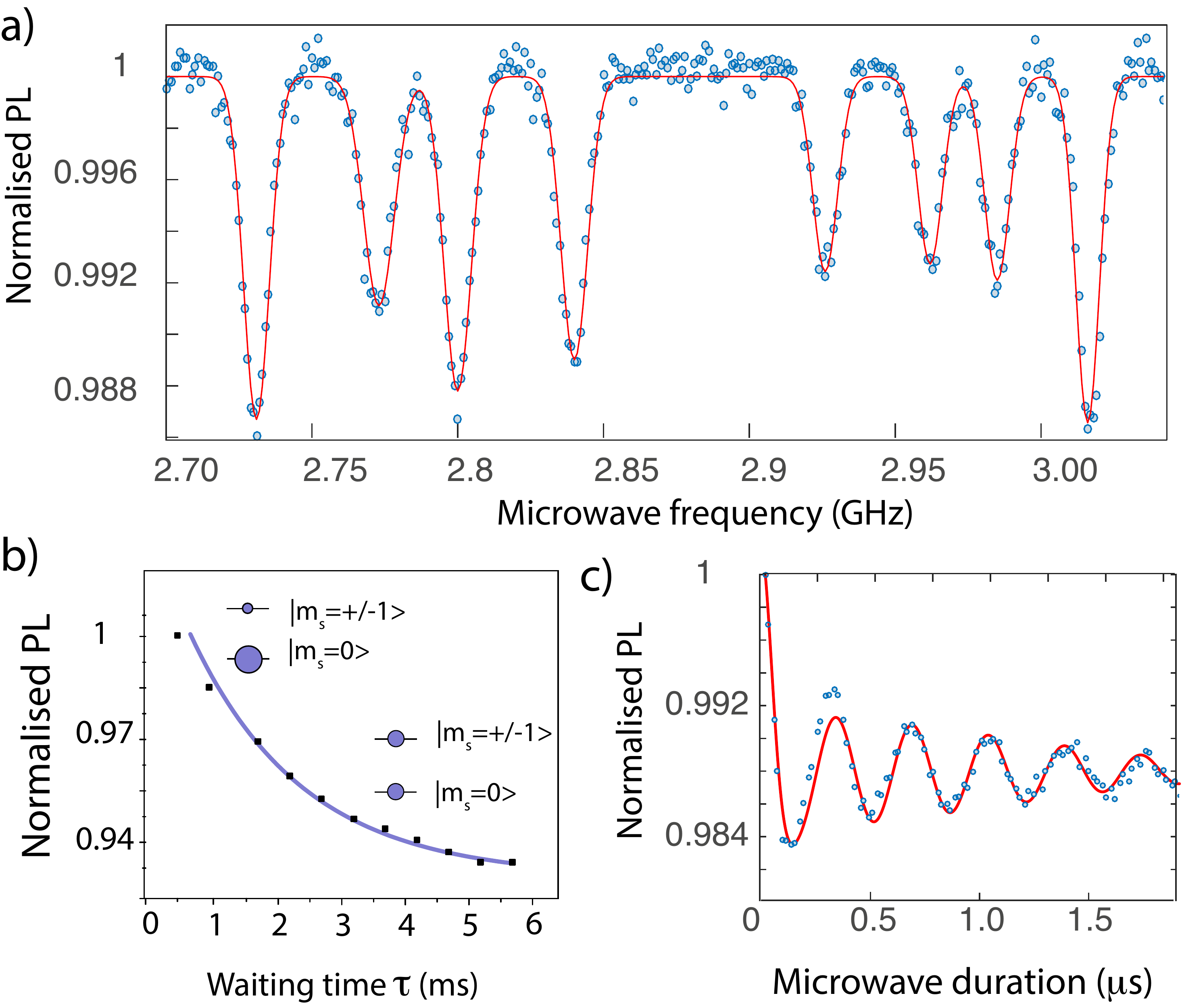}}}
\caption{a) Electronic spin resonance spectrum, b) longitudinal spin relaxation and c) Rabi oscillations from NV centers in a diamond monocrystal levitating in a Paul trap under ambient conditions.}\label{ESRRab}
\end{figure}

To realize spin coherent control, we apply a magnetic field that lifts the degeneracy between the state $|m_s=\pm 1\rangle$. As discussed in the SI (section II), crystal strain partially lifts the degeneracy between the $|m_s=\pm 1\rangle$ state and possibly also between the four orientations of the NV if anisotropic. Applying a magnetic field will thus ensure addressing of a more homogeneous class of NV centers. This will in turn reduce the ESR linewidth and improve the measured coherence time. 
Applying a magnetic field however means that the measurement must be realized on angularly stable particles (see SI, section II and III).
Fig.~\ref{ESRRab}-a) shows an ESR taken under atmospheric pressure in the presence of a magnetic field of about 50 G. 
Eight distinctive dips are observable, corresponding to the projection of the magnetic field over the four NV quantization axes inside the levitating diamond crystal.  
When comparing with typical ESR widths obtained with deposited diamonds (see section II of the SI), 
we conclude that the diamond does not rotate over the course of the measurement. This angular stability was interpreted in \cite{delord2016} to be due to a trapping mechanism akin to that of the center of mass when the shape of the diamond particles is asymmetrical (a SEM image of the diamonds is shown in section I of SI).
Here, the laser and microwave signals are just below the saturation of the transition, we then expect the ESR width to be determined mainly by the coupling to the diamond strain and coupling to impurities \cite{Dreau, Jensen}. Each ESR dip is indeed well approximated by a Gaussian function with an average width $\sigma$ of about 8 MHz. 
We thus expect $T_2^*$ times of about 50 ns. 


Another important quantity is the lifetime of the spin population in the $|m_s=0\rangle$ state, the so called $T_1$ time, which can be as long as milliseconds in bulk diamonds at room temperature.  
For opto-mechanical experiments involving spins, ideally the population lifetime time should equal half of the transverse coherence time for an optimum spin coupling to the motion coupling \cite{Bennett, Rabl}. 
It is thus important that the $T_1$ time is as large as possible, and possibly limited by electron to phonon processes, as can be the case in pure bulk diamonds \cite{Jarmola, BarGill}.
Fig.\ref{ESRRab}~-b) shows a measurement of the $T_1$ time of NV centers in the levitating diamond. 
The typical parameters of a sequence are detailed in the section II of the SI.
The evolution of the PL as a function of the waiting time shows exponential decay of the photoluminescence, indicating that the NV centers remains in the $|m_s=0\rangle$ state for more than 3 ms. 
Such a long $T_1$ highlights two features of this experiment : in our diamonds, the $T_1$ is close to typical to bulk values. Compared to nanodiamonds, it is thus not shortened by coupling of the NV spins to surface dangling bonds of paramagnetic impurities \cite{Tetienne}. Second, the Paul trap does not significantly modify the longitudinal spin properties of the NV centers.  
This offers the prospect of increasing $T_2^*$ towards millisecond long coherence times via dynamical decoupling technics \cite{Viola}.

\begin{figure}[ht!]
\centerline{\scalebox{0.26}{\includegraphics{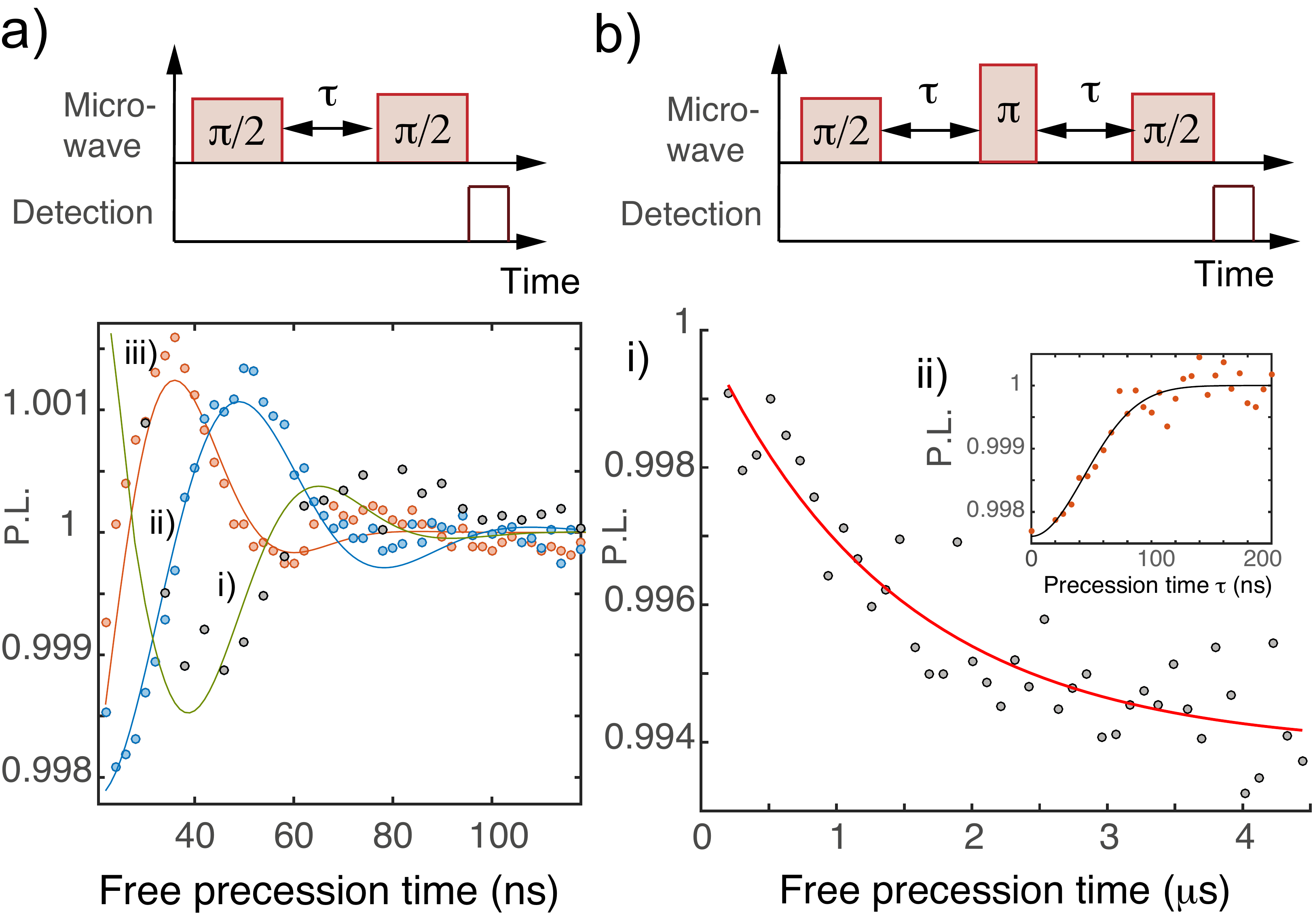}}}
\caption{a) Top : sequence used for measuring Ramsey oscillations with NV centers. Below : Ramsey oscillations from NV centers in a levitating diamond measured for three microwave detunings $\Delta/2\pi$ from the ESR line, under atmospheric conditions. b) Spin echo sequence and normalized photoluminescence as a function of precession time for the echo sequence (trace i)) and Ramsey measurements with a resonant microwave tone (trace ii)). }\label{Ramsey}
\end{figure}

We now demonstrate coherent control on many NV centers in the trapped diamond.
We chose the extremal $|m_s=0\rangle$ to $|m_s=+1\rangle$ ESR line and apply a sequence of microwave pulses with varying duration at this microwave frequency. Fig. \ref{ESRRab}-c) shows a plot of the normalized PL rate as a function of the microwave duration.
Rabi oscillations are observed for more than 1 $\mu$s.
The Rabi envelope decay is characteristic of the environmental noise spectrum \cite{Dobrovitski} (see section II of the SI). However, the observed damping time does not give direct access to the $T_2^*$ time. When the Rabi frequency is lower than the ESR width, less spectrally distinct NV centers' spin, and/or more classes of NV to nuclear spins couplings within the ESR profile are being excited, which effectively decreases the Rabi decay rate. The decay time can in fact be longer than the $T_2^*$ time and determined to a large extent by the employed microwave power \cite{Dobrovitski2, baibekov2011decay} (see also section II of the SI).

We now proceed with the demonstration of Ramsey interference with the NV spins in order to estimate the $T_2^*$ time of the NVs in the levitating diamond.
Ramsey interference is generally realized on two-level systems driven by two $\pi/2$ pulses separated by a time $\tau$, as depicted in Fig. \ref{Ramsey}-a).
To probe the coherence time, we detune the microwave and scan the time interval $\tau$ using levitating diamonds. Fig. \ref{Ramsey}-a) shows the change in the photoluminescence rate as a function of the free precession time $\tau$ for three different detunings from the central line and $\pi/2$ microwave pulse duration of 50 ns. 
Trace i) ii) and iii) correspond to detunings $\Delta/2\pi$ of 11, 15 and 20 MHz respectively. 
Since the three asymptotic PL values are different, the curves have been normalized by it for readability. 
A pronounced oscillation of the PL is observed as a function of the precession time and the precession period closely follows the inverse of the microwave detunings.
A Gaussian fit to the data yields decay times of $45\pm4$, $50\pm 5$ and $45\pm3$ ns respectively. 
Using a $T_2^*$ value of 47 ns, gives a corresponding ESR width of 9.4 MHz, very close to our measured ESR width value. 

To probe even further the capability of the Paul trap to preserve the electronic spin coherence, we now apply the spin-echo sequence depicted in Fig.~\ref{Ramsey}-b). In the HPHT diamond samples we use, temporal inhomogeneity resulting from nuclear spin impurities typically shift the energy of the NV centers' spins and affect their coherence on microsecond time scales (see section II of the SI), so applying a $\pi$ pulse between the two $\pi/2$ pulses can compensate for the associated spin dephasing. 
The result of the measurement is shown Fig.~\ref{Ramsey}-b)-i) where the normalized PL rate is plotted as a function of the precession time. The inset shows the corresponding resonant Ramsey curve.  The PL rate change is well approximated by a decaying exponential curve, from which we extract a decay rate of 1.4 $\mu$s  \cite{Dobrovitski2, Stanwix}.  Surprisingly, these values are typical to that observed with deposited diamonds (see SI, section II), so we conclude that the trap does not affect significantly the transverse spin coherence.

The above experiments were realized under ambient conditions. 
Typical opto-mechanical applications however require operation in the underdamped regime, that is in the regime where the collision rate with surrounding gas particles is smaller than the macro-motion frequency of the trapped diamond.  
As shown in the SI (section I), the underdamped regime is reached in the mbar range in our experiment.  To efficiently control the spins then, an important signature would be the presence of the four stable spin orientations in the ESR, which has not been observed so far in the underdamped regime.
Indeed, angular stability is much more challenging to achieve under vacuum, where the trapping frequencies are lower and any non-conservative force heats up the particle motion significantly (see SI, section II and VI).

\begin{figure}[ht!]
\centerline{\scalebox{0.27}{\includegraphics{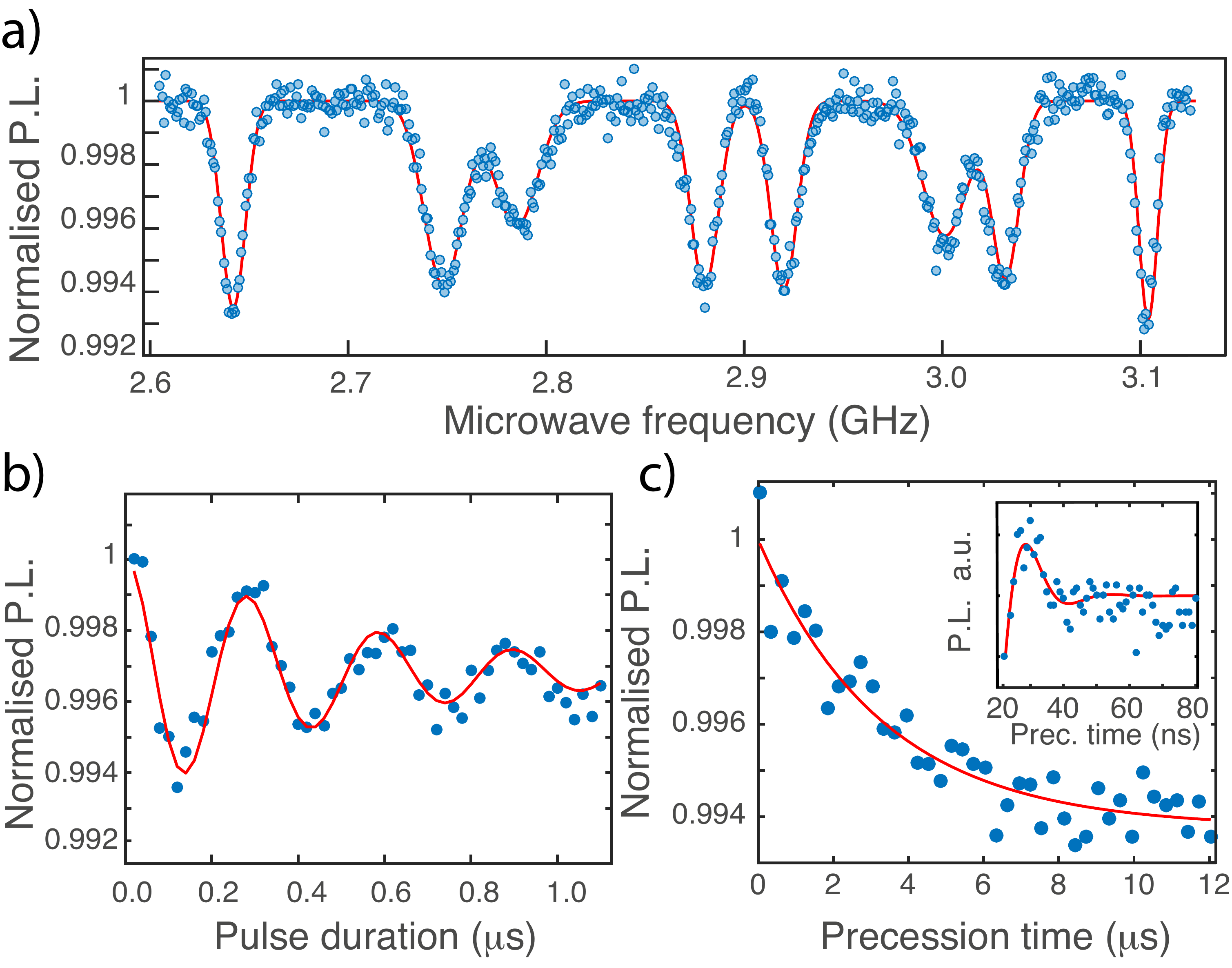}}}
\caption{Spin control of NV centers in diamonds levitating in the underdamped regime. a) Electronic spin resonance, b) Rabi oscillations and c) spin echo and Ramsey measurements (inset). All measurements are conducted under 1 mbar of vacuum pressure.}\label{RamVac}
\end{figure}

Fig \ref{RamVac}-a) shows an ESR spectrum taken under 1.5 mbar of vacuum pressure.
Although the ESR contrast is on average reduced due to the smaller employed microwave and laser powers (see SI - section II), the presence of eight distinctive Gaussian dips shows again that this diamond did not rotate significantly over the course of the measurement.  
The key ingredient to observing angular stability under vacuum was the fabrication of a small highly confining ring trap to counteract non-conservative forces, such as the laser radiation pressure torque on asymmetric particles \cite{delord2016} or residual micromotion.
This observation means that coherent driving on one NV orientation is possible.
Fig \ref{RamVac}-b) indeed shows Rabi oscillations taken at 1 mbars of vacuum pressure, where contrasted
coherent oscillations are observed with a decay times similar to that measured under atmospherical conditions. Fig.~\ref{RamVac}-c) shows spin echo and Ramsey (inset) measurements.
Ramsey measurements are taken at a detuning of 23 MHz. A clear variation of the PL is seen from 20 to 40 ns similar to the measurements done under atmospheric conditions. The decay time is estimated to be around $40\pm 10$ ns. Spin echoes show an exponential decay time of 3.3 $\mu$s. Again, there is no significant influence of the trap on the spin properties, even under this low vacuum level, where other levitating schemes suffer from heating that impacts the NV photo-physical properties \cite{Hoang, Neukirch2, vacuumESR, Pettit}. It is then likely that using purer samples, such as milled CVD diamonds \cite{Frangeskou}, or dynamical decoupling technics, will significantly increase the coherence time to milliseconds, very close to the oscillation frequency of our trapped diamonds.  

Our results will find direct use in the field of spin-opto-mechanics. 
A lot of effort is directed towards establishing a platform for coupling single spins to mechanical systems at the quantum level. Both tethered \cite{Rabl, Arcizet, Kolkowitz}
and untethered \cite{ma,yin, Kumar, delord2017strong} mechanical systems are being investigated. 
The latter makes use of single spins that are embedded in a moving particle in the presence of a fixed magnetic field. 
One promising way to establish strong coupling between the spin and mechanical degrees of freedom is to couple NV centers to
the rotational mode of a nanodiamond levitating in a Paul trap. The idea is to make use of the inherent quantization axis of the NV center to apply a torque to the whole nanodiamond. This can be done by coherently driving spin states dressed by a transverse magnetic field $B$ in the angular sideband resolved regime \cite{delord2017strong}. The coupling rate is proportional to the single phonon shift $\lambda_\phi=\gamma B \phi_0$, where $\gamma$ is the gyromagnetic ratio of the electron and $\phi_0=\sqrt{\hbar/2 I \omega_\phi}$ is the ground state extension of the angular mode. $I$ is the momentum of inertia, $\omega_\phi$ is the angular frequency of the considered rotational mode.   The strong coupling condition is attained when the spin-coupling rate to the rotational mode is larger than the decoherence time of both the spin and rotational mode. 

To achieve this goal, the coherence time of the NV center should not be impacted by the trap and the mean particle angle should be locked to a given position. Our observation of long spin echoes under vacuum together with angular stability thus confirms that combining a Paul trap with spins in diamond is a viable option for such a spin-opto-mechanical scheme.

To contemplate coupling the motion of the particle to the spin now, the confinement frequency (at present in the kHz range) should be increased. It can be dramatically improved by reducing the trap size, raising back the voltage under high enough vacuum (see section II, SI), adding endcaps electrodes. One can then also reduce the size of the particle to increase the charge to mass ratio.
Considering an prolate (aspect ratio 1:3) 180 nm diamond particle with a similar surface charge density than what we use here, in a 60 $\mu$m ring trap diameter and a peak voltage up to 3000 Volts, the frequency of the highest rotational mode $\omega_\phi$ is then expected to be around 100 kHz \cite{delord2017strong}, which requires a $T_2^*$ time of 10 $\mu$s to be in the sideband resolved regime will
enable entering the strong coupling regime where the single spin-phonon coupling rate to the motion is more than 60 kHz.\cite{delord2017strong}. Using lower vacuum pressures to minimize collisions with gaz particles and many NV centers \cite{yin} will then allow to be well within the strong coupling regime.

{\it Conclusion : } We demonstrate efficient coherent control of the spin of NV centers inside a levitating diamond.  
Spin echoes are employed to show that the surface charges and the high electric potential difference between the diamond and the ring Paul trap do not impact the coherence time of the spins on microseconds time scales. 
Furthermore, the NV centers are used as motional probes for the levitating diamond. We could indeed identify a regime where the trap strongly stabilizes the particle angle under vacuum against the angular micro-motion \cite{delord2017strong} and the laser radiation induced torque \cite{delord2016}.  These results establish the Paul trap as a robust platform for precision manipulation and detection of trapped macroscopic objects using embedded atom-like emitters.
Already now, our demonstration of angular stability already opens a clear path towards strong coupling to the rotational degree of freedom \cite{delord2017strong, yin}. 
A tantalizing prospect will also be to use dynamical decoupling technics \cite{Lange} to bring the NV center's coherence time close to the millisecond long $T_1$ time in order precisely measure the center of mass motion using magnetic field gradients.  This will offer clear prospects to experiments such as matter wave interferometry \cite{Scala,Wan}, quantum gravity sensing\cite{Bose2} strong coupling \cite{Rabl, Arcizet}, schrodinger cat state preparation \cite{yin2013optomechanics} which rely on the ability to maintain long coherence time for spin-state superpositions in a trapped object. 

\section*{Acknowledgements}
We would like to acknowledge fruitful discussions with Loic Rondin and Christophe Voisin.
This research has been partially funded by the French National Research Agency (ANR) through the projects SMEQUI and QUOVADIS (T-ERC).

\newpage

\begin{widetext}

\vspace{0.2in}
{\Large \hspace{2.13in}\textsc{Supplementary Material} }\\
\

\section{Experimental setup}

\subsection{Micro-diamonds injection and levitation under atmospheric pressure}

The experimental set-up is depicted in Fig.~\ref{setup2}-a). The trap and the objective are enclosed in a vacuum chamber.  The trap is a ring Paul-Straubel trap \cite{Straubel,Yu}. It consists in a small 15 $\mu$m thick gold-plated tungsten wire with an inner radius of 180 $\mu$m.
It is oriented so that the ring plane is perpendicular to the optical axis.

The diamonds that we used are in the form of mono-cristalline powders (MSY micron-diamond powder from Microdiamant AG) that did not undergo specific surface processing. They are produced by high-pressure, high-temperature synthesis (HPHT). They are injected using a small metallic tip that is dipped into the diamond powder and brought in the vicinity of the trap. Diamond sizes in the range of 8 to 12 microns are used in order to obtain a large number of NV centers and to ensure that single mono-crystals are injected in the trap. Using smaller particles (below 1 $\mu$m) often means that aggregates are injected. For particles with mean sizes above 12 $\mu$m, gravity dominates, so trapping is not effective.

With such micron-sized diamonds, we can operate the trap with a peak-to-peak voltage ranging from $V_{\rm ac}$=100~V to 4000~V at driving frequencies in the kHz range. Particles are generally injected under ambient conditions at $V_{\rm ac}$=4000 V and at a trap frequency of a few kHz.

A green laser is used both to monitor the particle position and to excite the NV centers inside the diamond. An acousto-optic modulator is used to switch on and off the laser beam.
The photoluminescence (PL) of the NV centers is collected using a confocal microscope with an aspheric lens ($f=8$ mm, N.A.$=0.5$). The PL signal is then filtered using a Notch filter centered at 532 nm and a long pass filter at 590 nm and can be directed either onto an avalanche photodiode or a spectrometer.
With a mW of laser excitation, we can collect around $10^6$ counts per second on the avalanche photodiode at atmospheric pressure. 

The micro-wave excitation was done via a Bias Tee, as depicted in Fig. \ref{setup}-a). Microwave powers of up to 0.5~W are used to excite the NV centers' electronic spins. 

\subsection{Levitation under vacuum conditions}

For the experiment realized under vacuum, the macro-motion frequency is lower.
Before lowering the pressure in the chamber, the voltage is indeed reduced to 600~V peak to peak to avoid plasmas that would otherwise appear at around 100 mbars in the chamber. 
Careful preselection of particles with a high charge to mass ratio is therefore performed at atmospheric pressures beforehand. To do so, we operate the trap at 4000~V, inject diamonds with a 10 $\mu$m diameter, and systematically measure the onset of instability. 

Another negative effect under vacuum is that the trap wire warms up significantly due to the microwave. It was seen that as the microwave power reaches 5 dBm in CW at around 1 mbars, that the particle would be shaken and sometimes even lost from the trap.  
Using a diamond that was deposited on the wire used in the trap, we could observe that the wire could reach 
500 K under CW microwave excitation at 5 dBm via NV center thermometry \cite{vacuumESR}.
This measurement was done at low laser powers, where we checked that the laser did not increase the diamond temperature above 300 K.
This means that convection can be at play, and increase the probability of losing the particle under vacuum, in the presence of the microwave. We thus, always reduce the microwave power when running ESR spectra under vacuum.

In Paul traps, stability is fulfilled if the trapping frequency is larger than than the secular frequency \cite{Pau90}. 
Measuring the trapping frequency at which instability takes places provides a means to estimate the charge to mass ratio. 
When this frequency is below 3 kHz at 4000~V, we estimated that the loss probability would be too high at 600~V and at low pressures so another particle is loaded. 
We also pre-select particles which contain as many NV centers as possible (we estimate the amount of NV centers to be about 1000 \cite{delord2016}) to get a large signal to noise ratio. 
The whole procedure typically requires 3 to 4 loading steps before a high enough charge to mass ratio and a dense enough concentration of NV centers is attained. 

Once a particle is trapped, for measurements under vacuum, the voltage and frequency are lowered in air, following an iso-$q$ curve, $q$ being the stability parameter of the trap at 1 bar \cite{Pau90}. Once 600~V is reached, the turbo-molecular pump is turned on and a valve is used to adjust the airflow in order not to loose the particle.

The green back reflected light can also be used to measure the particle motion along the ring plane. This measurement is typically done using a D-shaped mirror that cuts half of the light field mode and reflects it onto on one detector. The transmitted part is detected and the signals from the two detectors (PDB210A, from Thorlabs) are subtracted and sent onto a spectrum analyser (Keysight N9020A MXA). We performed a measurement with a ring that was 700 $\mu$m in diameter.
Under 0.5 mbars of vacuum pressure, the power density of the subtracted signal displays two Lorentzian peaks at 119 Hz and 161 Hz. The frequency position of these two signals is seen to shift when the trapping frequency $\Omega$ is modified. 
We checked that these macro-motional frequencies scale like the inverse of $\Omega$ as expected. These two frequencies are compatible with the two first center of mass modes of motion in the trap, their frequency difference reflecting the slight asymmetry of the ring trap. The ring trap is perpendicular to the optical axis, so the secular motion along the $z$ direction, which should appear at around 280 Hz is not detectable.
The smaller ring that is used for the coherent transients presented in the manuscript implies that the oscillator is well within the underdamped regime at 1 mbar of vacuum pressure.

\begin{figure}[ht!]
\centerline{\scalebox{0.25}{\includegraphics{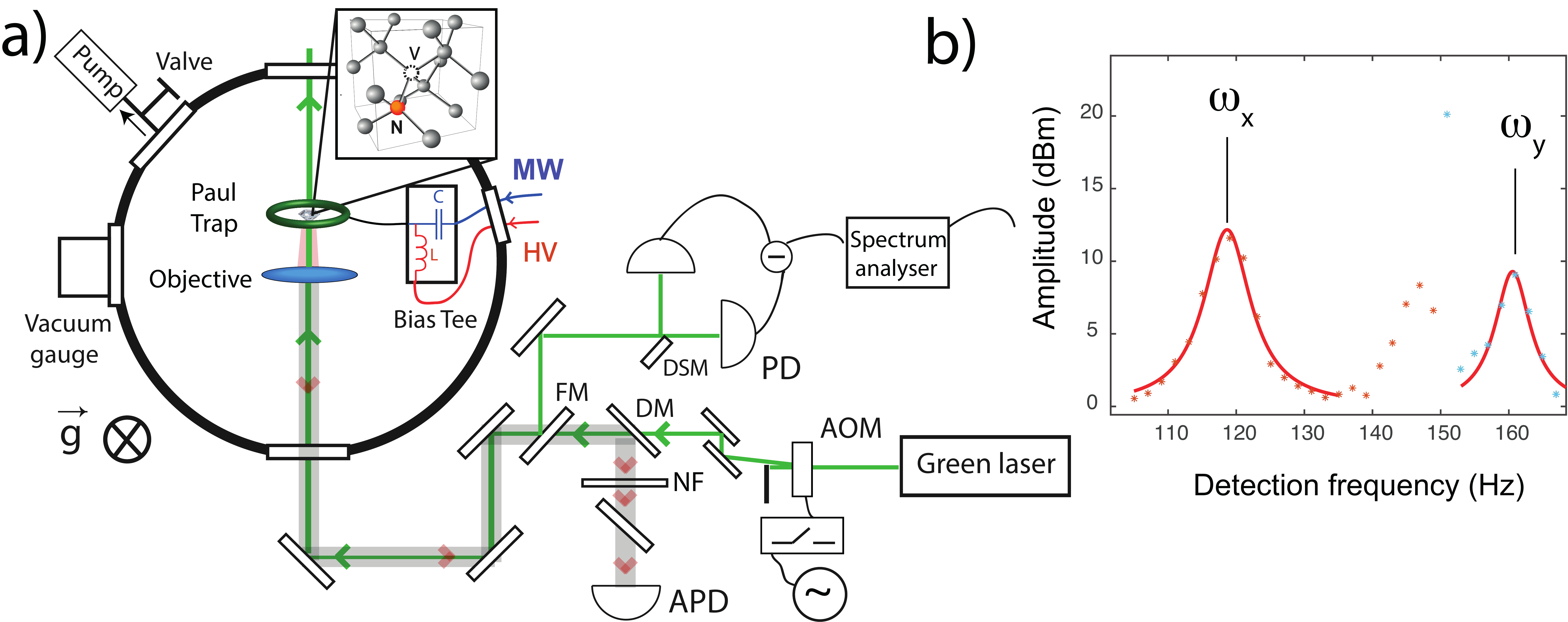}}}
\caption{a) Schematics of the experimental apparatus. The Paul trap, objective, and Bias Tee are enclosed in a vacuum chamber. 
A green laser is focused onto a diamond levitating in the Paul trap.  The photoluminescence from the NV centers in the trapped diamond is collected by the same objective and measured either on an avalanche photodiode (APD) or on a spectrometer. 
DM= dichroic mirror, NF=Notch filter centered at 532 nm. FM=flipping mirror. PD=Photodetector, HV=High Voltage, MW= Microwave. DSM= D-shaped mirror. AOM= Acousto-Optic-Modulator b) First two center of mass modes at 300 V of trap voltage and 1.12 kHz of trap frequency at a vacuum pressure of 0.5 mbars. The noise in between corresponds to a multiple of 50 Hz noise together with artefacts that do not depend upon the trap parameters
}\label{setup2}
\end{figure}

\subsection{Diamond inhomogeneities}

The shape of the diamonds was found to be highly inhomogeneous, as can be seen in the SEM image Fig. \ref{MEB}.
As shown in \citep{delord2016}, this has strong consequences both for angular stability and sensitivity to the green laser.
We also found that the NV density is highly inhomogeneous from one diamond to another \citep{delord2016}. Besides the differing count rates, this also changes the green laser power required to polarize the NV spins \cite{Jensen}. 

As we discuss next, another consequence of these differing shapes is that the level of strain for each NV orientation is different. 
Such a difference from one diamond to another means that we must perform a careful analysis of the deposited diamond properties to compare them with the levitating diamond ones.

\begin{figure}[ht!]
\centerline{\scalebox{0.5}{\includegraphics{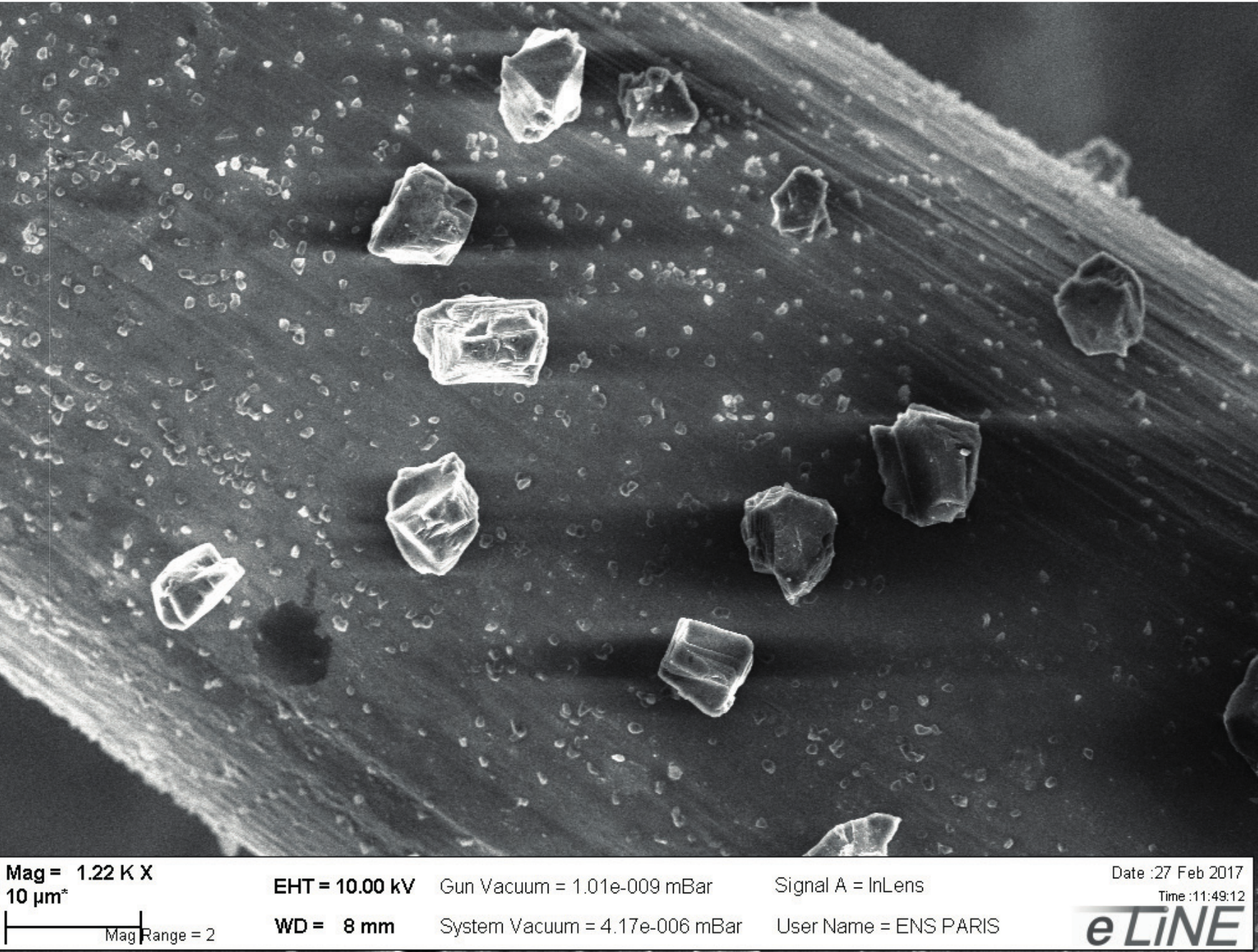}}}
\caption{Scanning Electron Microscopy (SEM) image of micro-diamonds deposited on the metallic tip used for loading.}\label{MEB}
\end{figure}

\section{Spin characterization of NV centers in deposited micro-diamonds}

Let us know characterize the spin of NV centers in the micro-diamonds.
The NV center level structure is shown in Fig. 3. a). 
The electronic spin is manipulated using microwave tones in the ground state $^3A_2$. 

In order to assess the influence of the Paul trap on the spin properties of embedded NV centers or on the ESR signals we performed measurements on micro-diamonds that are deposited on a quartz coverslip.
Here we present ESR measurements and estimate the electronic spin $T_1$ and $T_2$ times using Rabi and Ramsey oscillations. Note that the conditions and measurement parameters were kept similar to the experiments carried out with levitated micro-diamonds.

It was not possible to compare the ESR when the same particle is deposited and when it is levitating, so instead we perform a characterization of the spin properties of NV centers for deposited diamonds to check if the spin properties are similar.

\subsection{Electron spin resonance (ESR)}

\begin{figure}[ht!]
\centerline{\scalebox{0.3}{\includegraphics{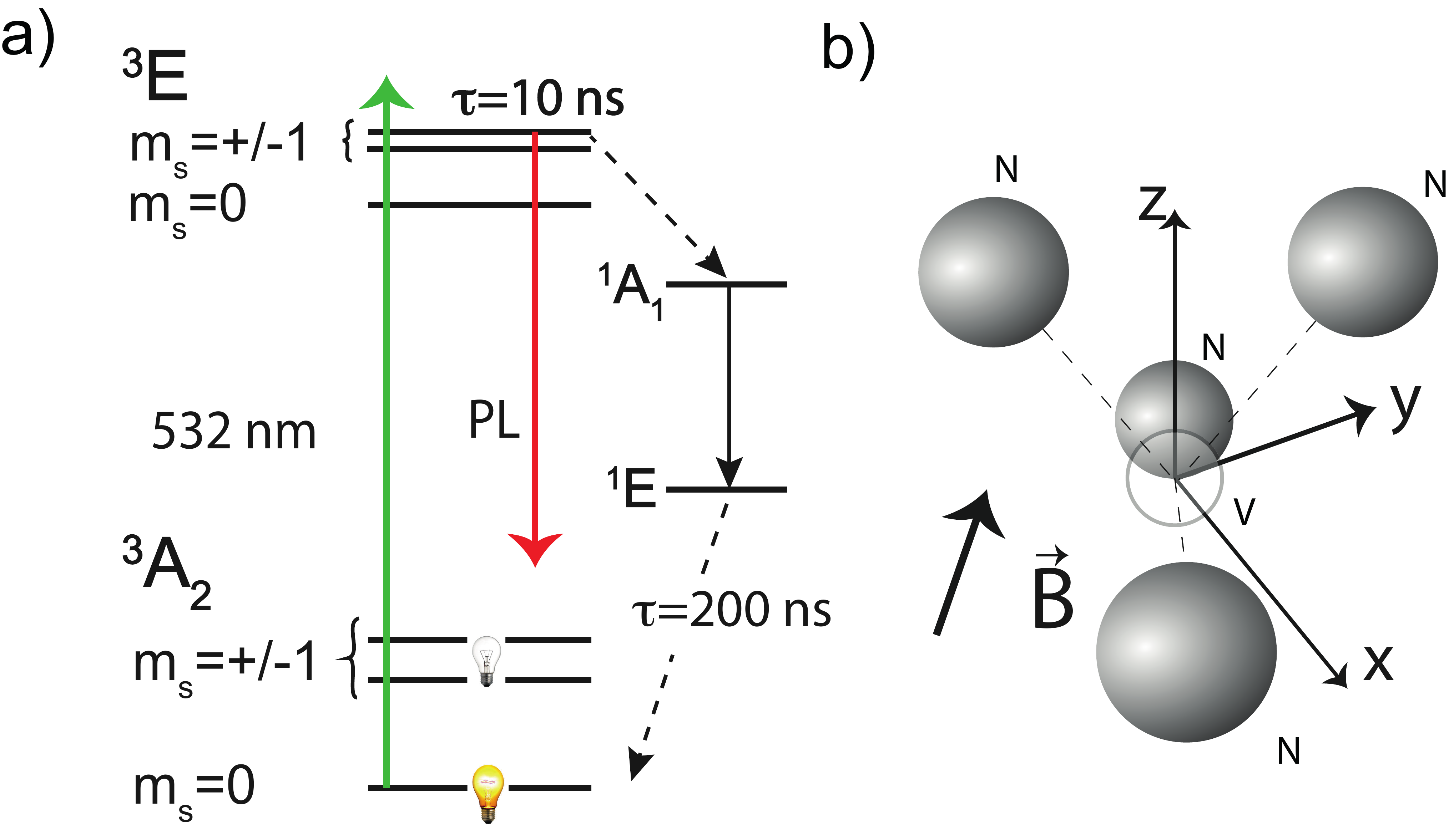}}}
\caption{a) NV level structure. b) One possible orientation of the B field with respect to the diamond cristal, with four different projections of the NV axes on the B-field. }\label{levels}
\end{figure}

\subsubsection{ESR without an externally applied magnetic field}

Fig. 4-a) shows a measurement of the photoluminescence of NV centers inside two different diamond particles in the absence of applied magnetic field.
The two traces i) and ii) were measured using the same laser and microwave intensities. The microwave power has been chosen so that the ESR lines do not undergo power broadening, simply by reducing the power until the linewidth does not change.
As in the case of the levitating diamond whose ESR is displayed in Fig. 2 of the main text, these ESRs show two lobes. Here they are separated by 6 and 14 MHz for trace i) and ii) respectively. In the manuscript, it was 10 MHz for the ESR  taken under ambient conditions. 
This splitting is caused by crystal strain. 
As can be seen there is more than 50\% difference between the strain induced splittings from one diamond to another. 

For each orientation the NV Hamiltonian reads 
$$H=D S_s^2+ E(S_x^2-S_y^2).$$
Where $D/2\pi=2.87$ GHz and $E$ models the mixing between the $m_s=0$ and $m_s=\pm 1$ levels due to strain. 
In our case, double Gaussian functions does not fit our data, most likely because the four NV orientations do not undergo the same strain. In this case, the ESR signal would thus be a superposition of four double Gaussian curves, each with a different $E$ value. 

Importantly, we noticed no significant difference in the distribution of strain values (centered around $2E$=10 MHz with a range of values from 5 to 15 MHz) inside and outside the trap. 

\subsubsection{ESR with an externally applied magnetic field}

Applying a magnetic field lifts the degeneracy between the NV magnetic spin states and narrows down the width of the ESR. 
As depicted in Fig.3b, four orientations of the NV in the diamond monocrystal are possible. This means four projections of the magnetic field on the four quantization axes of the NV centers are possible. Here, for this diamond orientation a spin transition $m_s=0$ to $m_s=-1$ was isolated spectrally at 2.613 GHz.

In the manuscript, we present ESR under atmospheric condition with -4dBm of microwave power, below the saturation of the transition.
We can observe clearly the four orientations. The ESR contrast varies between 0.7\% to 1.2 \% depending on the microwave polarisation and on the transverse static magnetic field strength with respect to each NV axis \cite{Tetienne2}. 

For the experiments done under vacuum the microwave power have been lowered to $-10$dBm to avoid convection from the trap that disturbs the particle motion. 
In the continuous wave regime, different microwave powers lead to a different ESR width and a change in the contrast \cite{Dreau}.
This is illustrated in Fig. \ref{ESR_pow}-b), where two ESR spectra measured around the NV spin transition at 2.613 GHz is displayed for two different microwave powers at a laser power of 2 mW under an applied magnetic field of around 50 G.
Fig. \ref{ESR_pow}b)-i) and -ii) shows the normalized photoluminescence as a function of the microwave frequency with microwave powers of 11~dBm and 1~dBm respectively. 
A Gaussian fit to the data gives a contrast of 2.4\% and a width of 11 MHz for trace i) and a contrast of 1.5\% and a width $\sigma=7.9$ MHz for trace ii).
The reduction of the contrast at low microwave powers is due to a competition between the laser power and the microwave. The ESR width tends to the $T_2^*$ value as the microwave signal is reduced.
Integrating a longer time under a lower microwave power can then be used to estimate the $T_2^*$ time. 
Another option is to perform the experiment under pulsed excitation using $\pi$ pulses that are longer than $T_2^*$\cite{Dreau}. 
The $T_2^*$ is measured using Ramsey oscillations in the article, which avoids complications related to such power broadening. 

Under low microwave saturation, the ESR width is a result of the magnetic dipole coupling between the NV centers and randomly distributed nuclear spins of
paramagnetic impurities (P1 centers), $^{13}$C isotopes, fields caused by local
electric fields, and the interaction between the NV electronic spins with their nitrogen nuclear spin. These couplings are similar to an inhomogeneous broadening as at a given time, each NV center coupled to its neighboring spins has a different energy.
Far from saturation, the resulting ESR inhomogeneous linewidth should indeed be a Gaussian \cite{Maze2, Zhao, Dobrovitski} and related to the $T_2^*$ time via the formula $\sigma=2 \ln(2)/(\pi T_2^*)$.

\begin{figure}[ht!]
\centerline{\scalebox{0.4}{\includegraphics{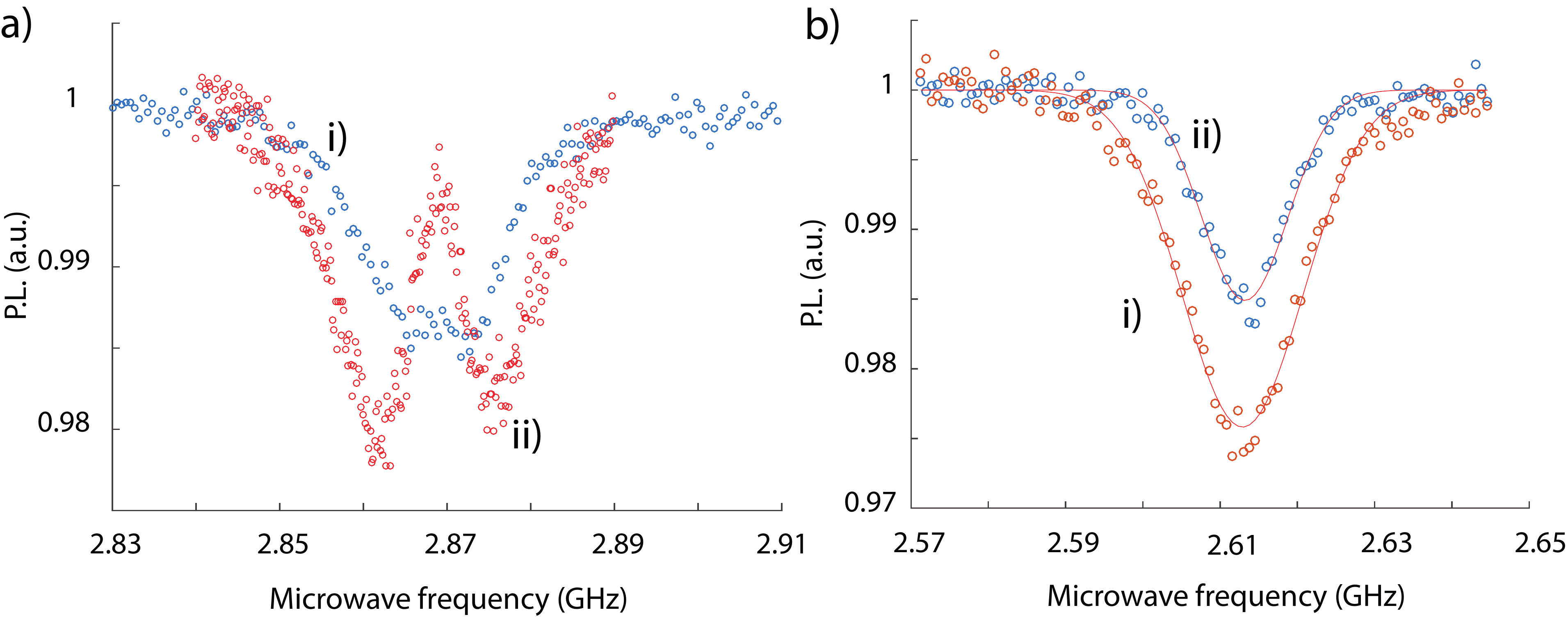}}}
\caption{a) ESR measured without externally applied magnetic field on two different deposited micro-diamonds. b) ESR signals measured at 11 dBm and 1 dBm for trace i) and ii) respectively, using deposited diamonds with an externally applied magnetic field. 
The plain lines shows a Gaussian fit to the data.
}\label{ESR_pow}
\end{figure}

\subsection{Longitudinal spin relaxation}

Let us first present measurements of the longitudinal spin relaxation, that is the so-called $T_1$ time of the electronic spin. 
To estimate the $T_1$ time in our samples we use the pulse sequence depicted in Fig. \ref{T_1}-a).
A 15 microseconds, 1 mW green laser pulse first polarises the NV center's electronic spins by using an intersystem crossing in the excited state, as depicted in Fig\ref{levels}-a).
An identical second laser pulse is used to excite the NV center and to collect the photo-emission. A PCI card (PCI-express National instrument) counts the avalanche photodiode (APD SPCM-ARQ-15 from Perkin-Elmer) signal during 10 microseconds. The amplitude of the measured PL is proportional to the probability of being in the ground $|m_s=0\rangle$ state.
It thus measures how fast the NV spins relax back to the thermal equilibrium. The time necessary to polarize the spin state depends on the optical power used and can be estimated looking at the contrast whereas the window used for measuring the PL is obtained by optimizing the signal of noise ratio. Typically the measurement is repeated about 10000 times per point, depending on the PL rate.

Fig. \ref{T_1}-b) shows a typical result for deposited diamonds. A decay time $T_1=3.2 \pm 0.2 \rm ms$ was inferred from an exponential fit to the data. 
Our measured $T_1$ values actually range from 1 ms to 10 ms limited possibly by coupling of the electronic spin to the phonon bath of the diamond crystal or by nearby impurities \cite{Jarmola}.

The $T_1$ time is long enough to attribute the decay of the Ramsey oscillations in the micro-diamonds to a decay of the coherence only as we will see.
 
\begin{figure}[ht!]
\centerline{\scalebox{0.5}{\includegraphics{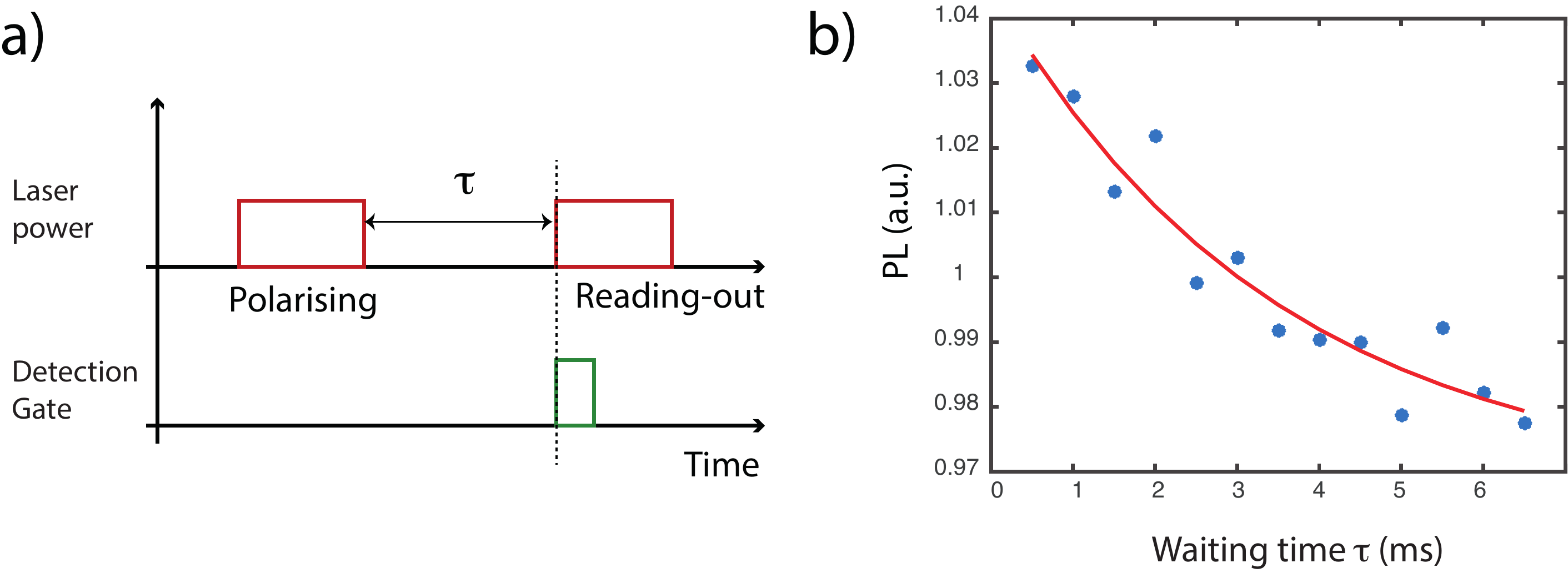}}}
\caption{a) Shows the pulse sequence used for measuring the $T_1$ time of NV centers. b) Photoluminescence as a function of the dark time between the two laser pulses.  An exponential fit gives $T_1=3.2 \pm 0.2 $ms.}\label{T_1}
\end{figure}

We also measured the laser induced depolarization rate of the $|m_s=\pm1\rangle$ states.
This measurement was done using the sequence shown in Fig.\ref{Polarisation}-a).
A microwave signal is applied for $1\mu$s on one of the ESR to prepare a mixed state with equal population in the $|m_s=\pm 1\rangle$ state and $|m_s=0\rangle$.
The laser is kept on at all times. 
The PL rate is measured a time $\tau$ after the applied microwave, and determines the amount of population that is being transferred to the $|m_s=0\rangle$ state by the laser.
The result of the measurements is shown in Fig.\ref{Polarisation}-b).
An exponential fit (solid line) reveals that it takes 5 $\mu$s to depolarize the spins prepared in the $|m_s=\pm1\rangle$ states. It also means that it takes more than $20~\mu$s to efficiently polarize them to the $|m_s=0\rangle$ state with about 1 mW of laser power.

\begin{figure}[ht!]
\centerline{\scalebox{0.4}{\includegraphics{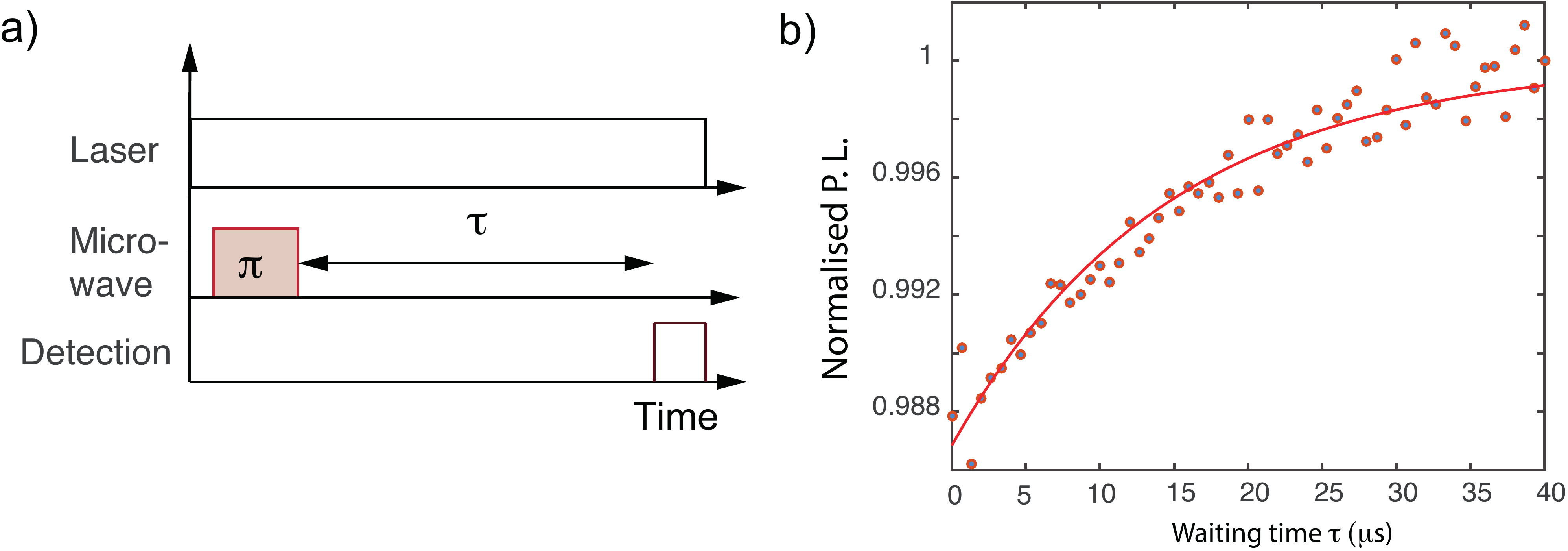}}}
\caption{a) Sequence used for measuring the green laser polarisation rate of the NV center into the $m_s=0$ state. b) P.L. rate as a function of waiting time, showing that it takes more than 20 microseconds to efficiently optically polarise the NV center's spin.}\label{Polarisation}
\end{figure}

\subsection{Transverse relaxation}

We now present measurements of the Rabi and Ramsey oscillations of the electronic spins of NV centers in deposited diamonds. 

The microwave is generated by a Rohde \& Schwarz SMB100A RF generator and sent to a microwave RF switch (ZASWA-2-50-DR+ from Minicircuit).
We then use a card (PulseBlaster from SpinCore Technologies, Inc.) to generate TTL pulses which control the RF switch and amplify the output signal (ZHL-5W 422 from Minicircuit) before sending it to the bias-tee or directly to an antenna. 

\begin{figure}[ht!]
\centerline{\scalebox{0.4}{\includegraphics{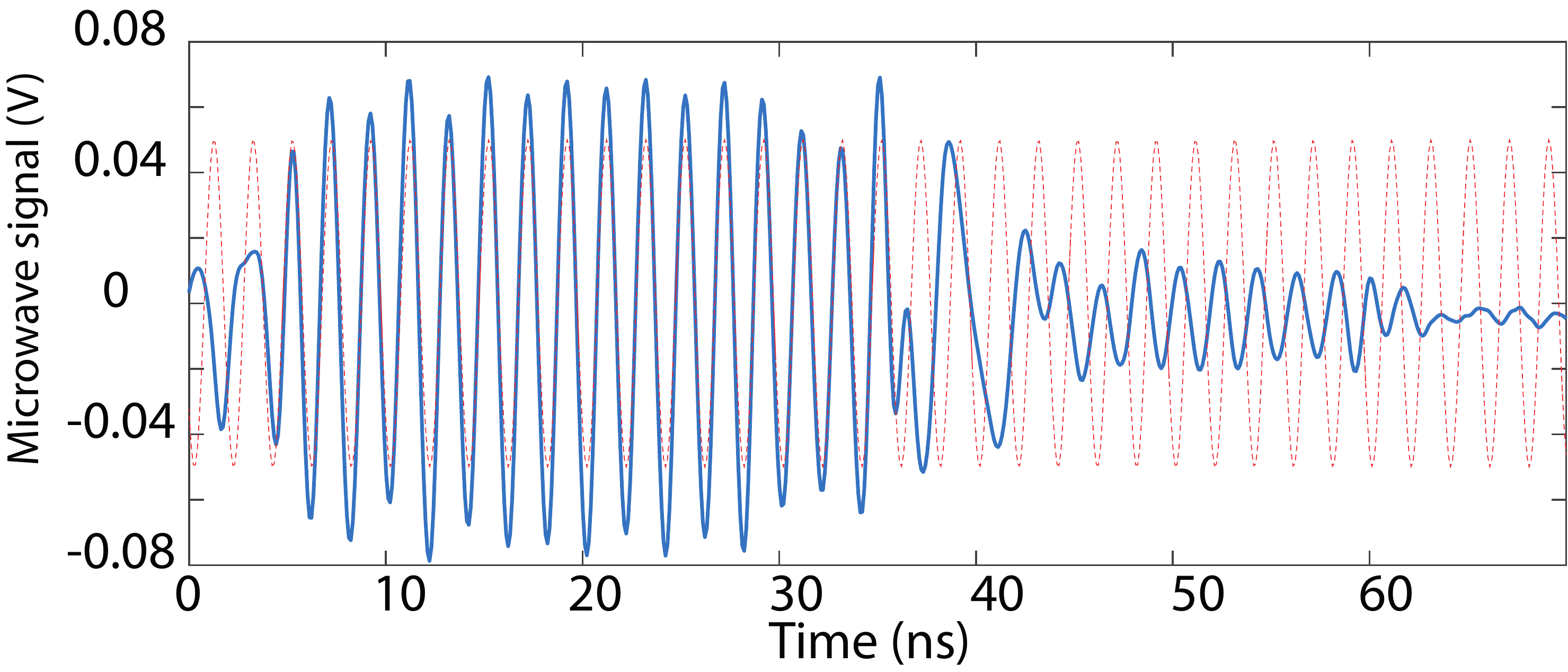}}}
\caption{Microwave signal detected at a frequency 500 MHz on an oscilloscope. A ring-down is observed from 40 ns to 60 ns. A dashed red line shows a reference signal at the same frequency, to highlight the phase shift between the signal and the after pulse.}\label{afterpulse}
\end{figure}

Fig. \ref{afterpulse} shows a time trace of the microwave signal amplitude measured after the amplifier at a frequency of 500 MHz on an oscilloscope. The TTL signal sent to the microwave switch is 35 ns.
A ring-down that lasts 20 ns was observed. 
This ring down amplitude is 20\% that of the generated signal and, as shown by the reference dashed line, is out of phase with it for these measurements. Simulations on homogeneous samples suggest that using such a pulse for Ramsey interferometry or Rabi oscillations significantly impacts the spin evolution below 20 ns.
Modeling the dynamics of the many emitters that are all coupled differently to a bath of nuclear spins in the presence of such a ring-down is beyond the scope of the present manuscript. We thus consider the Ramsey and Rabi dynamics above 20 ns to discard this ring down contribution.

\begin{figure}[ht!]
\centerline{\scalebox{0.5}{\includegraphics{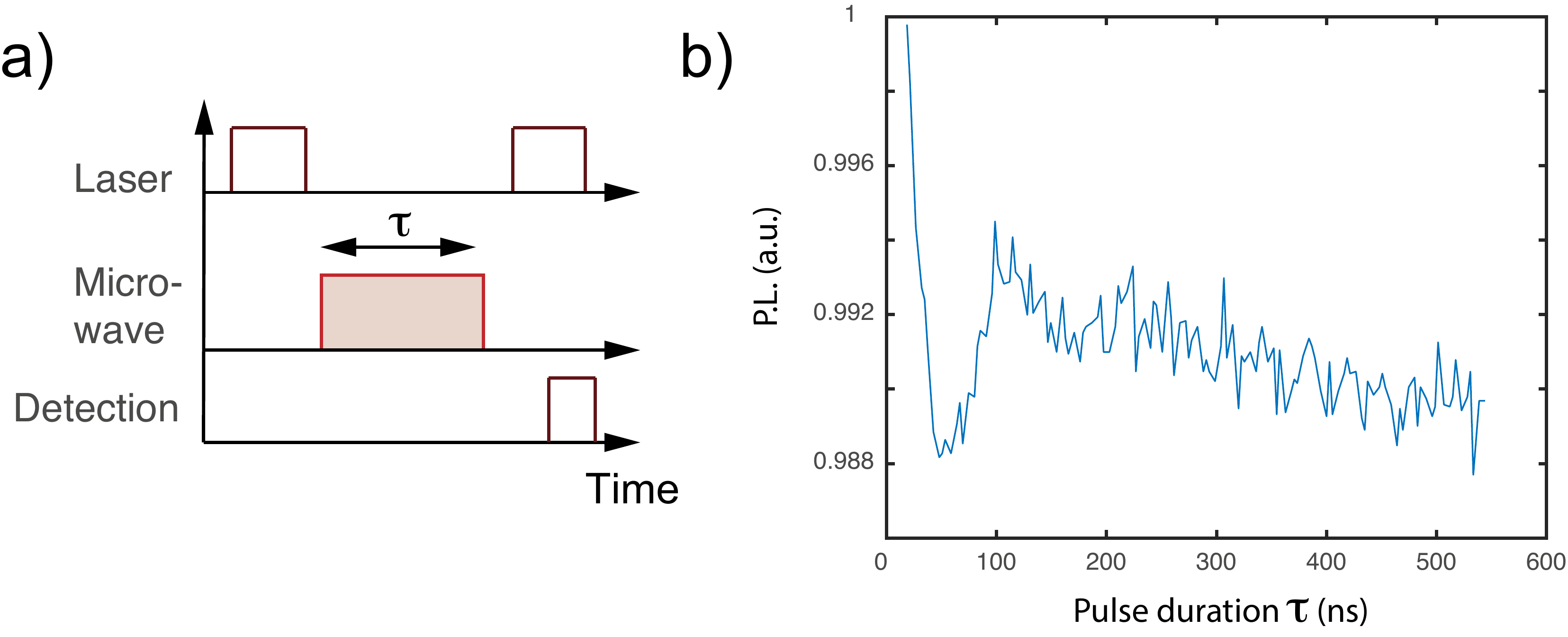}}}
\caption{a) Sequence used for measuring Rabi oscillations with a pulsed green laser.
b) Rabi oscillation from NV centers in deposited diamonds without externally applied magnetic field. This measurement was taken at a microwave frequency tuned to one lobe of the ESR shown Fig.4 a), trace ii). }\label{Rab3}
\end{figure}

\begin{figure}[ht!]
\centerline{\scalebox{0.4}{\includegraphics{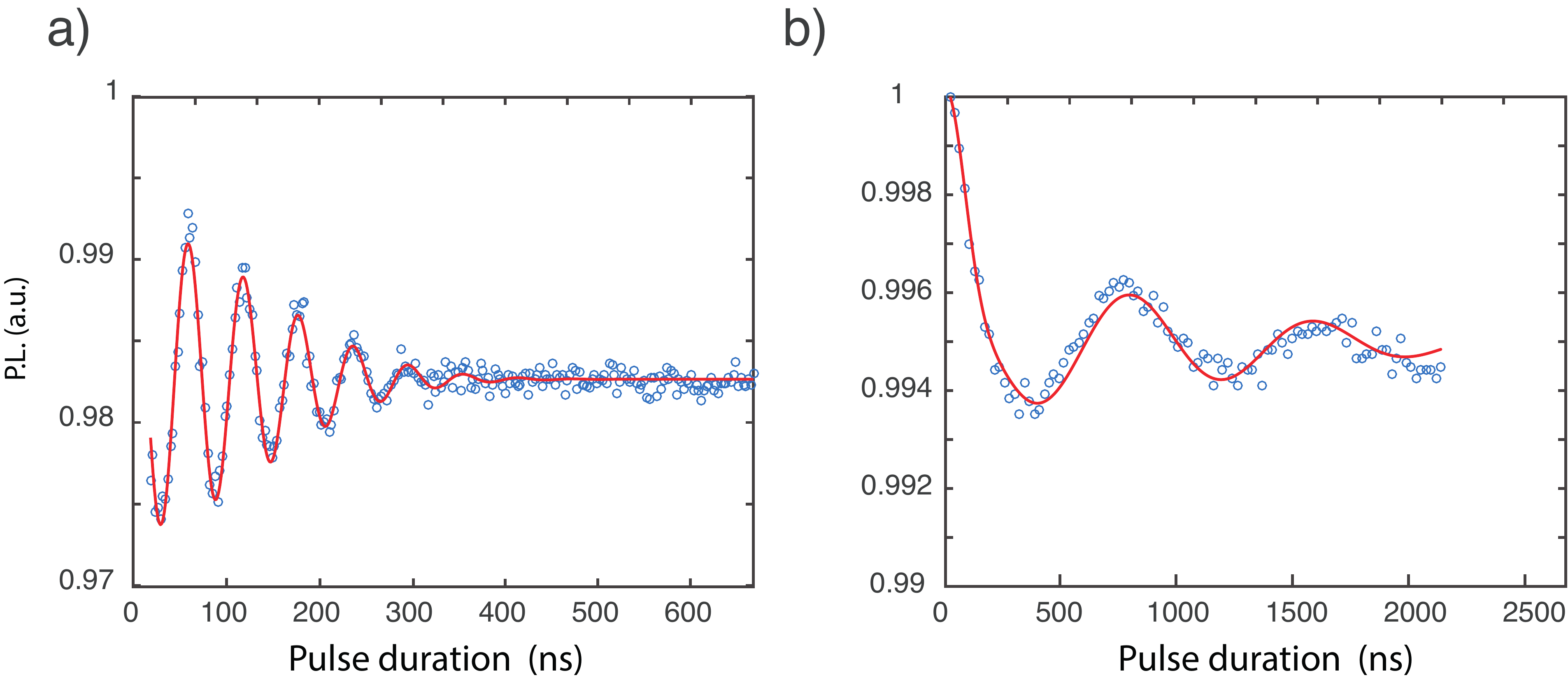}}}
\caption{Rabi oscillations from NV centers in deposited diamonds for two different microwave powers in the presence of an externally applied magnetic field. All parameters are the same apart from the two microwave powers.}\label{Rab}
\end{figure}

\begin{figure}[ht!]
\centerline{\scalebox{0.4}{\includegraphics{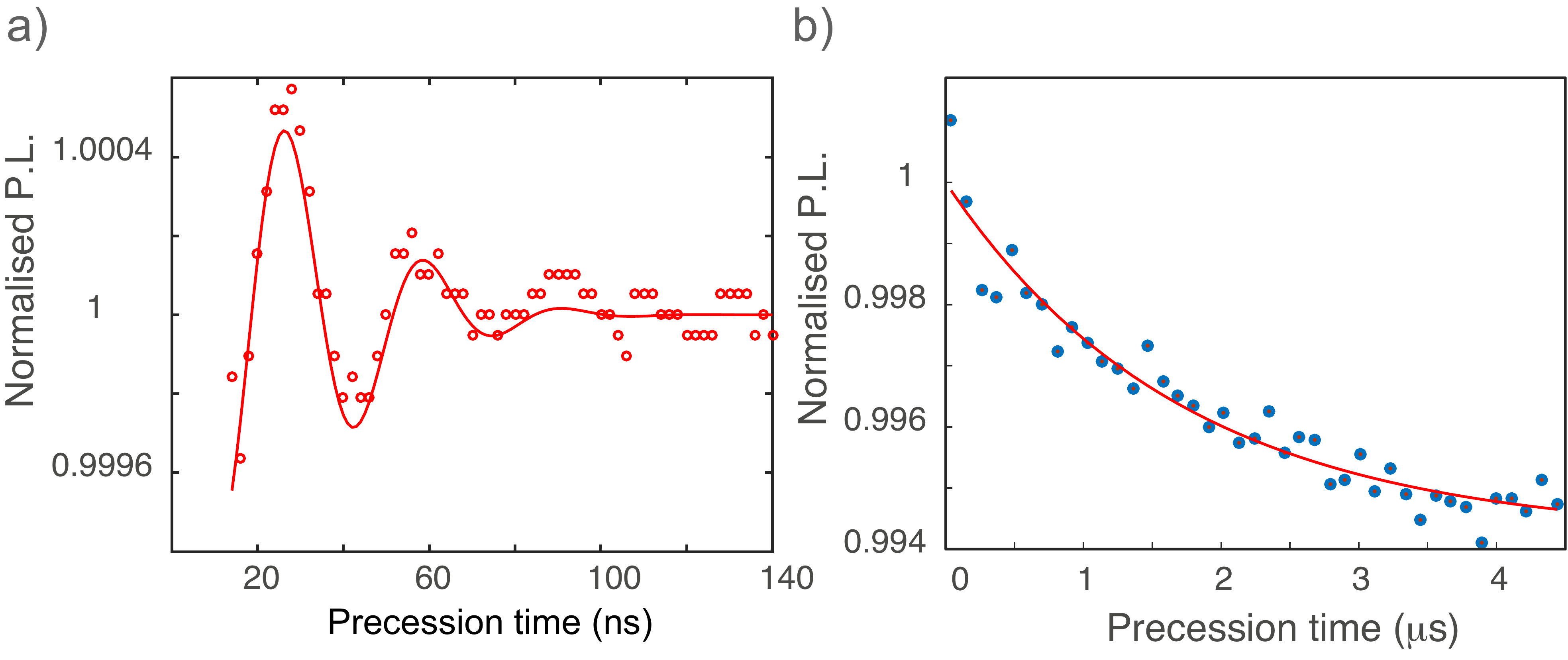}}}
\caption{a) Ramsey oscillations. b) Spin echoes from NV centers in deposited diamonds.}\label{Ram}
\end{figure}

To measure Rabi oscillations, we bring a microwave antenna that is similar in shape and size to the ring that we use for trapping close to the deposited micro-diamonds. 

For most of the results presented in the manuscript the laser was kept on. 
As shown in Fig. \ref{Polarisation}-b), the laser does not significantly reduce the population lifetime of the NV centers electronic spin when kept on for a time below 5 $\mu$s. Most of the measurements were thus performed with the laser on during the whole sequence.  As we will see, the echo measurements show a coherence time of 3 $\mu$s at most and the Rabi oscillations are performs for time of up to $2\mu s$ so the laser can be kept on for all the coherent measurements as the spin population dynamics will be much faster than the polarization rate. One can however expect the laser to affect the measured $T_2$ time differently than the measured $T_1$. To verify this, we performed extra measurements by running the sequence shown Fig. \ref{Rab3} at the same time as a sequence where the laser is kept on at all times 
on several particles inside and outside the trap when performing Rabi, Ramsey or spin echo measurements.
These measurements gave the same result making us confident that we can keep the laser on for measurements performed on time scales below $5$ $\mu$s.

As already discussed, without magnetic field, the ESR width is larger due to the difference in strain for the four NV orientations. 
We thus expect a faster Rabi decay in the absence of B field. 
Fig. \ref{Rab3} shows a measurement of Rabi oscillations without externally applied magnetic field, with a frequency tuned to the left side lobe of the ESR of Fig. 4a- trace ii), that is at 2.865 MHz. Although this Rabi curve is the most contrasted that was observed without magnetic field, the Rabi oscillations last only 100 $ns$.

We now approach a permanent magnet close to the diamond and drive one orientation of the spin ensemble at a frequency 3.158 GHz in Fig. \ref{Rab}. 
Fig. \ref{Rab}-a) and b) show the NV photoluminescence as a function of the micro-wave pulse duration for two different microwave powers. In a) we use 20 dBm, while in b) we use -5 dBm.

For all the Rabi oscillations fits, we use the function
\begin{equation}f(t)=1-C_1 (1- (C_3 e^{-t^2/\tau_1^2}+C_5 e^{-t^2/\tau_2^2})\cos \Omega_R t).\label{functon_fit} \end{equation}
More involved theory that include the inhomogeneous broadening (both spatial and due to the nuclear spins)
can be used \cite{Dobrovitski, Dobrovitski2}.
Due to the above mentioned after-pulse issue and the large parameter space to cover, getting a good agreement between this theory and our data is a difficult task. The theory of \cite{Dobrovitski2} could only fit 1/3 of our Rabi oscillations.
We thus simply intend to give estimates that we can use to compare the results inside and outside the trap. 
The theory of  \cite{Dobrovitski, Dobrovitski2} predicts a dependence of the Rabi decay with the microwave power for a single spin coupled to a nuclear spin bath. 
This non-trivial damping dependence on the Rabi frequency is however not specific to NV centers but is also observed with spin ensembles in solid states with a strong inhomogeneous broadening \cite{boscaino1993non,baibekov2011decay}. A simple description of the phenomena is that, as the Rabi frequency is increased, classes of magnetic dipolar coupling rates to the NV that are far from the center of the ESR line will contribute more. As is clear from Fig. 9, Rabi oscillations can indeed be much shorter than the inverse of the ESR width when $\Omega_R/(2\pi)> \sigma$. The different resulting couplings will lead to faster dephasing which will result in a higher decay rate, although with a higher asymptotic contrast.

For the first Rabi measurements, the fit gives the following values :  
$C_1 =0.017,
 \Omega_R =0.10,
C_3 =-5.500\times10^{-4},
\tau_1 =179,
C_5 =0.52,
\tau_2 =192,$
in nanoseconds units. It yields Rabi oscillations periods of 60 ns$\pm$4 ns.
The decay is about 190 ns and so it can be modeled by a single Gaussian decay to a good approximation. 

For trace b), we get the following values for the fit 
$C_1 = 0.0048,
 \Omega_R =0.008,
C_3 =0.7778,
\tau_1 =176,
C_5 =0.2548,
\tau_2 =1565.$
Two decay constants are thus non negligible here. The short term evolution decays with a time constant of 176 ns while the long term one is given by 1.565 $\mu$s.
These results shows that the dynamics depends strongly on the employed microwave power (see next section). 

Ramsey fringes are more precise for measuring the $T_2^*$ when using short intense $\pi/2$ pulses. The sequence used in the paper is shown in Fig. \ref{Ram}-a).
Fig. \ref{Ram}-b) shows the resulting Ramsey oscillation with deposited diamonds and with a microwave detuning of 25 MHz from the ESR transition. 
The microwave pulse duration is 50 ns. This Ramsey signal gives a $T_2^*$ of 45 ns, very close to the values expected from the Gaussian ESR width whereas even the shortest measured Rabi decay rate is not in accordance with the ESR width.

Spin echoes are shown in Fig. \ref{Ram}-b)
In these measurements, we apply two $\pi/2$ pulses separated by a $\pi$ pulse and change the free precession time.  
This procedure rephases the NV centers' spins evolution on the Bloch sphere when their evolution takes place on time scales longer than the time interval between two $\pi/2$ pulses.
The decay is very well approximated by a Gaussian decay. It yields a decay time of 1.65 $\mu$s.
We typically measured values that range between 1 to 3 $\mu$s.
with profiles that can be approximated by Lorentzian or Gaussian profiles depending on the impurity and NV center concentration \cite{Dobrovitski2}. 

\section{Rabi oscillations with NV centers in levitating diamonds}

In this section, we present Rabi oscillations that have been obtained with a diamond levitating under atmospheric conditions. 
As highlighted in section I-A, different diamond particles may have different properties. 

We show here that similar Rabi response with respect to the microwave power is obtained when the diamonds are levitating. Fig. \ref{Rab_atm} displays Rabi oscillations from NV centers' spins levitating in a micro-diamond for three different Rabi frequencies. 
Using the same fit as described above, we find long time decays of about 1.5 $\mu$s, 1.3 $\mu$s and 0.8 $\mu$s when using 11, 15 and 20 dBm of microwave powers for traces a), b) and c) respectively. Note that 20 dBm were used for the plot of Fig. 3-b) in the core of the manuscript. 
where we extract two decay times by fitting the Rabi decays to two Gaussian envelopes : 81 $\pm$ 5 ns  and 1183 ns $\pm$ 10 ns.
Similar dependence of the decay time are thus observed when using a levitating diamond.
As long as the diamond does not rotate over the course of the measurement we do not expect additional decay. This is indeed the case in these experiments, for which the ESR signals a stable angle.  

\begin{figure}[ht!]
\centerline{\scalebox{0.25}{\includegraphics{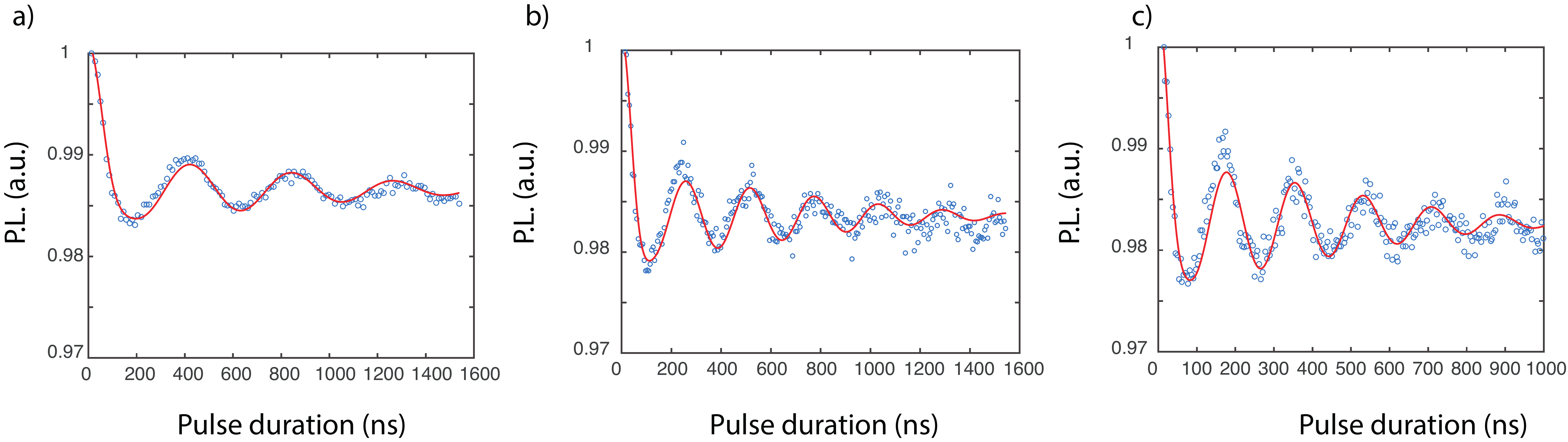}}}
\caption{Rabi oscillations from NV centers' spins levitating in a micro-diamond with microwaves powers of 10, 15 and 20 dBm for traces a), b) and c) respectively.}\label{Rab_atm}
\end{figure}

\begin{table}[ht!]
\centerline{\scalebox{1}{\begin{tabular}{|l||c|c|c|c|c|c|}
\hline
{\bf Power} & $C_1$ & $ \Omega_R$  & $C_3$ & $\tau_1$ & $C_5$ & $\tau_2$  \\
\hline
10 dBm & 0.013 & 11.1 & 0.84 & 133 & 0.21 & 1536\\
\hline
15 dBm 	& 0.016 & 18 & 0.86 & 95 & 0.23 & 1292 \\
\hline
20 dBm	& 0.017 & 26 & 0.90 & 60 & 0.31 & 849 \\
\hline
\end{tabular}}}
\caption{
Parameters used for the fits of the Rabi oscillations shown in Fig. \ref{Rab_atm} using the function from equation \ref{functon_fit}. The Rabi frequency is expressed in rad/$\mu s^{-1}$ and the decay times are in $ns$.}\label{rabi_table}
\end{table}

Table \ref{rabi_table} lists the parameters used for fitting the Rabi oscillations at the three different Rabi frequencies, for this particle (different than in the core of the the manuscript). These three regimes yield drastically different decay constants. In the three scenarios the Rabi frequency $\Omega_R/(2\pi)$ can be three times smaller, two times smaller and almost equal to the inhomogeneous width $\sigma$. 

\section{Levitating diamonds under vacuum}

Under vacuum condition, any heat inside the levitating diamonds can be kept for much longer times so the temperature in the diamond matrix can rise considerably. This is due to the absorption of the green laser by impurities \cite{millen2014nanoscale,Rahman,Frangeskou}. This alters the NV properties notably by increasing non-radiative transitions but also by shifting the ESR line frequencies. This effect in turn allows efficient read-out of the internal temperature \cite{Hoang, vacuumESR, Pettit}.
Another effect is convection due to heat from the trap which implies that the microwave power has to be reduced during the ESR. We use -10 dBm, for Fig. 5-a)
of the manuscript and 15 dBm for Fig. 5-b) and c). 
A similar fit than for the plot of Fig. 3-b) in the manuscript yields two decay times of 70 ns and 1.1 $\mu$s, similar to what was measured under atmospheric conditions.

\begin{figure}[ht!]
\centerline{\scalebox{0.4}{\includegraphics{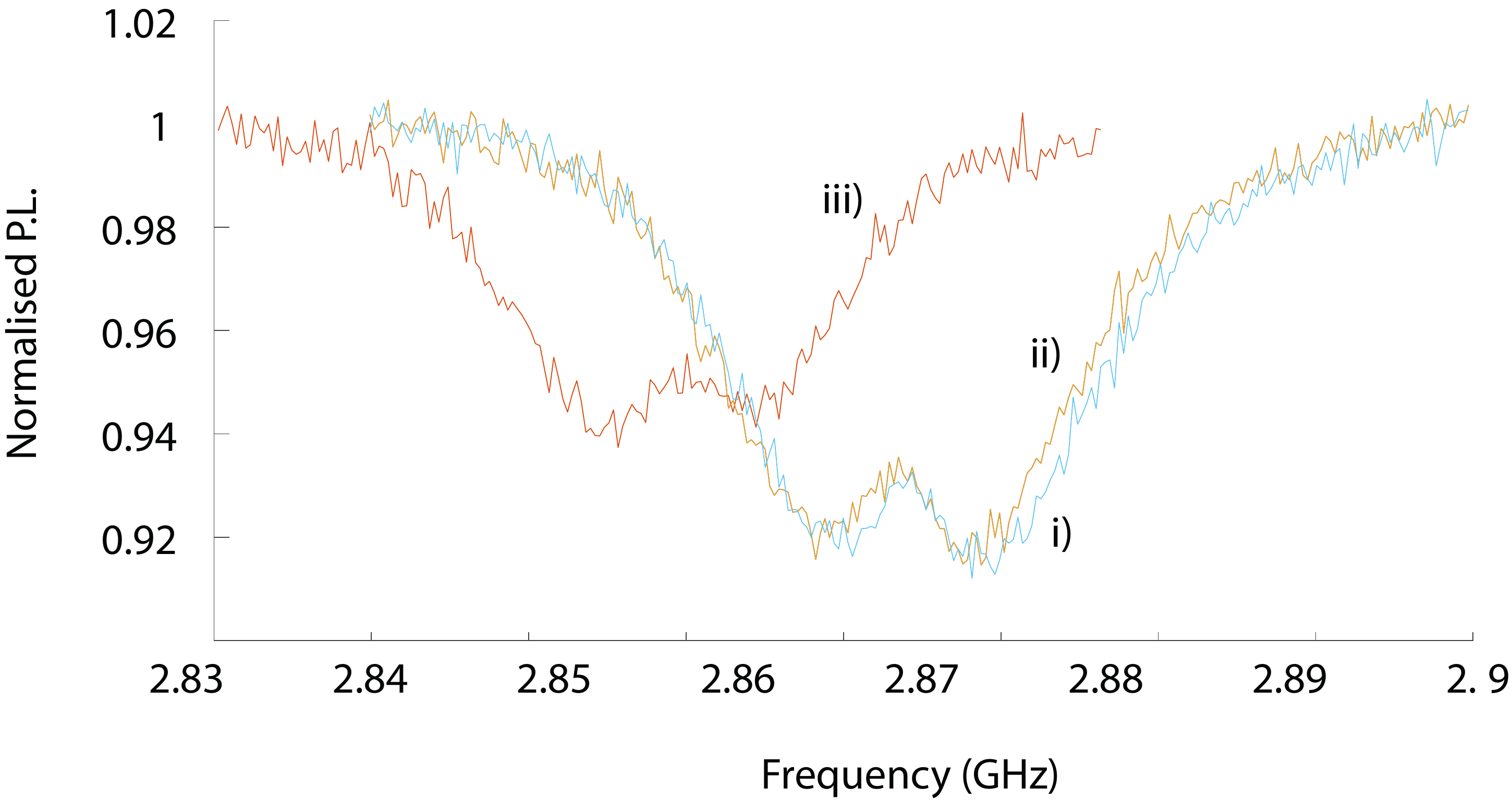}}}
\caption{Electronic spin resonance spectra without magnetic fields for different vacuum pressure levels. 
i) P= 1mbar. ii) P=8.9$\times10^{-2}$ mbars iii) P=3.8$\times10^{-2}$ mbars with a laser power of 100 $\mu$W.}\label{Temp}
\end{figure}

We can extract temperature values from the central position of the ESR minima.
A pronounced shift of the zero-field splitting (ZFS) can be observed in the ESR shown in Fig.~\ref{Temp}.
Since the lattice extension changes with temperature, the factor $D$ depends on temperature. 
$D$ was accurately described by a third-order polynomial between 300 K and 700 K in \cite{Toyli}
\begin{eqnarray}
D(T)=a_0+a_1 T^1 + a_2 T^2 +a_3 T^3,
\end{eqnarray}
where $a_0=2.8697$ GHz,
$a_1=9.7\times 10^{-5}$ GHz/K, and $a_2=-3.7\times 10^{-7}$GHz/K$^2$
which yields shifts of about 80 kHz/K close to 300 K.
Using these values, we can deduce the temperature of the levitating diamond from the frequency shift of the ZFS.

Absorption of light in the material and subsequent heating can be attributed to nitrogen defects inside the diamond since nitrogen is naturally present in commercially available HPHT diamond powders \cite{Rahman}.  
In the experiment performed in Fig. \ref{Temp}, heating is mitigated by using a minimum of 0.5 mbars and not more than 200 $\mu$W of laser power. This ensures that no shift in the ZFS takes place and that the diamond does not heat up significantly.

\section{Influence of the diamond rotation on Rabi oscillations}

We here discuss the decay of the Rabi oscillations from NV centers in a rotating diamond, with an emphasis on Rabi frequency averaging.
For an angularly stable diamond, the probability $P(t)$ of being in the excited $|m_s=\pm 1\rangle$ state at times shorter than $T_2^*$ for a given class of NV centers, is given by the simple Rabi formula 
$$P(t)=\frac{1}{2}(1-\cos(\Omega t)),$$
where $\Omega$ is the Rabi frequency, related to the projection of the microwave polarization onto the NV quantization axis.
If now the diamond rotates in a plane that is not perpendicular the microwave polarization direction, the microwave coupling to the NV will change as a function of the diamond angle. The Rabi frequency can thus take many values.
For a rotation that is slow compared to the smallest Rabi frequency, averaging of the Rabi oscillations over the frequencies range $\Delta \Omega=\Omega_{\rm max}-\Omega_{\rm min}$
gives
$$\overline{P(t)}=\frac{1}{2\Delta \Omega} \int_{\Omega_{\rm min}}^{\Omega_{\rm max}}~d\Omega~(1-\cos(\Omega t))
=\frac{1}{2}\Big[1-\cos({\overline{\Omega}t})~{\rm sinc} \big(\frac{\Delta\Omega t}{2}\big)\Big]$$
where $\overline{\Omega}=(\Omega_{\rm max}+\Omega_{\rm min})/2$ is the averaged Rabi frequency. Here, the decay of the Rabi oscillations will take place at a rate on the order of the covered Rabi frequency range $\Delta \Omega$. Depending on the diamond rotation axis with respect to the microwave, this effect can make the observation of long coherent transients more difficult. A similar effect will take place in the presence of an externally applied magnetic field.
Next, we discuss reasons for possible rotations of our particles.

\section{Center of mass temperature under vacuum conditions and angular stability}

Under vacuum conditions, the absence of dissipation also impacts the center of mass and rotational degrees of freedom and their associated temperatures may rise significantly in the presence of a heat source. Normally, the residual gas both heats up and cools down the levitating particle leading to brownian motion and friction forces respectively.\\
When the underdamped regime is reached however, the radiation pressure force is the dominant heat source. There are two ways for radiation pressure to heat up the particle : through photon shot noise or because of its non conservative property.
The photon shot noise induces a small heating that can only be measured at very low temperature \citep{jain2016direct}, conversely the heating mechanism due to the non conservative property of the radiation pressure is non-negligible, notably because of the high temperature. This heating can be understood simply : because the force is not conservative, there can be non-zero work added within one cycle of the particle oscillation \cite{roichman2008influence} and a flow of energy can be injected into the system. 
The competition between this heating and the friction from the residual gas yields an out of equilibrium steady state, with a temperature that depends upon the laser field intensity and shape. Although this effect has for now mainly been discussed and studied in the case of optical tweezers \cite{roichman2008influence,wu2009direct,li2013millikelvin}, a Paul trap with an added excitation beam qualitatively reproduces similar conditions possibly with a more complex motion with superimposed micro-motion.

Without an additional cooling mechanism, such as parametric feedback cooling, the center of mass temperature of the particle can therefore rise when the pressure is lowered and the particle can eventually be lost. If we now consider the angular degree of freedom of an angularly stable particle, we can expect that a similar behavior can occur~: when the pressure is lowered, the temperature of the angular degrees of freedom rises and the angular stability will be degraded until the harmonic confinement of the angle is lost \cite{delord2017strong}.\\

In this article, Rabi and Ramsey oscillation are carried out under a homogeneous magnetic field. A high external temperature of the levitating diamond can then broaden the observed ESR lines because each NV orientation will have a time-varying spin energy as the NV axes rotate compared to the magnetic field. This effect could imply a slight drop of the contrast for the Rabi and Ramsey oscillation under higher vacuum since the oscillations would be averaged over different detuning and frequencies, as described in section V.
In practice, at around 30 mbar of vacuum pressure and if the laser power is kept in the mW range as for the atmospheric pressure experiments, the particle starts to rotate and the laser power is usually lowered to several hundreds of micro-watts in order to stabilize it. A quantitative estimation of the role of the laser and the influence of micro-motion on the rotational degree of freedom will be left to further studies.

\end{widetext}

\end{document}